\documentclass{article}
\usepackage[dvips]{graphicx}
\usepackage{fullpage,}

\newcommand\bea{\begin{eqnarray}}
\newcommand\eea{\end{eqnarray}}
\newcommand\ga{{\gamma}}
\newcommand\bfA{{\mathbf A}}
\newcommand\bfB{{\mathbf B}}
\newcommand\bfl{{\mathbf l}}
\newcommand\bfr{{\mathbf r}}
\newcommand\bfZ{{\mathbf Z}}

\begin{document}


\begin{flushright}
\vskip-0.75cm
UdeM-GPP-TH-05-129\\
\end{flushright}
\bigskip

\begin{center}
{\LARGE Comment on ``On the Electric Charge Quantization from the
  Aharonov-Bohm Potential''}\\
\bigskip\bigskip
R.~MacKenzie, H.~Paquette, J.~Pinel, P.-L.~Roussel\\
\it Laboratoire Ren\'e-J.-A.-L\'evesque, Universit\'e de Montr\'eal\\
C.P. 6128, Succ. Centre-ville, Montr\'eal, QC H3C 3J7
\end{center}

\bigskip
\begin{abstract}
In the paper \cite{barhel}, Barone and Halayel-Neto (BH)
claim that charge quantization in quantum mechanics
can be proven without the need for the existence
of magnetic monopoles. In this paper it is argued that their
claim is untrue.
\end{abstract}
\bigskip


In the paper \cite{barhel}, Barone and Halayel-Neto (BH)
claim that charge quantization in quantum mechanics
can be proven without the need for the existence
of magnetic monopoles. The argument relies on a re-analysis of the
Aharonov-Bohm (AB)
effect, as follows. The magnetic field of an infinitely long
solenoid of radius $R$ lying along the $z$ axis is
\bea\label{bsol}
\bfB_{\rm sol} (\bfr)=\hat z B \,\Theta(R-\rho),
\eea
where $\Theta(x)$ is the Heaviside step function, equal to 0 or 1 for
$x$ negative or positive, respectively.
This is described in \cite{barhel} by the vector potential
\bea\label{vecpot}
\bfA(\bfr)=\hat\phi\left\{{B\rho\over2}\Theta(R-\rho)+
{\ga\over2\rho}\Theta(\rho-R)\right\},
\eea
where $\gamma$ is, a priori, a free constant.

Were $\ga=BR^2/2$ (the standard choice in the literature, written
$\ga_{\rm AB}$ in \cite{barhel}), this vector potential indeed
describes $\bfB_{\rm sol}$; with any other value of $\ga$ there is (as
noted in \cite{barhel}) in addition a magnetic field localized at
$\rho=R$ whose integrated flux is $\Phi'=2\pi\ga-\Phi_{\rm sol}$,
where $\Phi_{\rm sol}=B\pi R^2$ is the flux of the solenoid. BH
describe the cylinder $\rho=R$ as
``non-physical,'' possibly because it is a ``singularity region for
the field strength, since there is a surface charge density,''
which apparently justifies their use of a vector
potential describing a magnetic flux (as measured by $\int \bfA\cdot
d\bfl$ integrated around a curve encircling
the solenoid) unequal to $\Phi_{\rm sol}$. BH parameterize their
choice of vector potential by $\kappa\equiv \ga-\ga_{\rm AB}$, so that
$\Phi'=2\pi\kappa$.

It is then argued that the solenoid will exhibit the usual AB effect
(with the correct magnetic flux, $\Phi_{\rm sol}$)
for particles of charge $q_i$ only if
$q_i\kappa \in \bfZ$, which is to be viewed as a condition
on $\kappa$. In order for this to occur for a particle of
charge $q_1$, we must therefore have $\kappa=n_1/q_1$, where $n_1$ is
some integer. A second charged particle, of charge $q_2$, must then
obey $q_2\kappa=n_2$, $n_2\in \bfZ$, so that
\bea\label{chquant}
q_2={n_2\over n_1}q_1.
\eea
This is the charge quantization condition as derived by BH.

Our objections can be put into two categories. Firstly, and most
bluntly, the vector
potential (\ref{vecpot}) is simply wrong. Rather than declaring the
position of the solenoid to be ``unphysical'', since indeed $B_{\rm
  sol}$ is discontinuous there, and so ignoring the flux in that
region, one should recognize that (\ref{bsol}) is the field of an
ideal solenoid made of infinitely thin wires infinitely close to one
another. A better treatment of this ``singularity region'' would be to
smooth out the discontinuity in
the magnetic field (equivalent to considering
finite-thickness wires) so that it is equal to $B$ inside a certain
radius and zero outside a second slightly larger radius, with a
smooth interpolation between these two values.
Such a magnetic field can easily be
described unambiguously (up to gauge transformation) by a smooth
vector
potential without invoking fictitious magnetic fluxes which must
subsequently
be rendered unobservable by insisting on a charge quantization
condition.

But let us put aside this argument and examine the reasoning used in
 \cite{barhel}, and its consequences. Starting with the consequences,
 their main result (\ref{chquant}) is an exceedingly weak charge
 quantization 
 condition, indeed. It merely states that all charges must be related to
 one another by rational factors. To illustrate the weakness of
 (\ref{chquant}), note that while it must be admitted that
 charges $e$ and $\sqrt{2} e$ are not compatible, $e$ and 1.4142136$e$
 are. (The reader may replace $\sqrt2$ by her/his favourite irrational
 number, and 1.4142136 by an arbitrarily accurate rational
 approximation to it.)

What of the reasoning itself? Essentially, the authors let the value
of one charge determine the possible values of $\kappa$, which then
determines the possible values of all other charges in the
problem, by insisting that the flux $\Phi'$ causes no AB effect for
any particle.

(While somewhat peripheral to the discussion at hand,
note that this seems to select one charge
[the one which determines the allowed
values of $\kappa$ -- that
is, the allowed strengths of the fictitious
delta-function magnetic field which can be added] as having a
special role. Indeed, the allowed strengths of the
fictitious magnetic field would be
different if this initial charge were $e$ or 1.4142136$e$.
One might argue
that this is not a serious problem, since, after all, it is
arranged that the
fictitious magnetic field is unobservable. The more important point,
perhaps, is that if the initial charge were $e$, then 1.4142136$e$ is
allowed, and vice versa. Essentially,
the choice of with which charge one begins is immaterial in so far as
charge quantization is concerned,
because (\ref{chquant}) is symmetric.)

However, why not turn the argument around and argue that the
additional delta-function field must be constrained by insisting that,
whatever charges exist, this additional field must yield no AB effect?
In other words, rather than having one charge constrain $\kappa$ and
then having the allowed values of $\kappa$ constrain subsequent
charges, why 
not let the pre-existing charges (which obey no a priori quantization
condition) constrain $\kappa$? For instance,
if $e$ and 1.4142136$e$ existed, certain magnetic fields could
exist without having any observable effects. (To be specific,
writing 1.4142136=$m/n$
with $m,n$ relatively prime integers, one finds that
the allowed values of $\kappa$ 
are integer multiples of $n/e$.) However, let us apply this same
reasoning
to a situation where charges $e$ and $\sqrt2e$ exist. In order for
neither charge to experience an AB effect, it is necessary that
$\kappa$ is an integer multiple of {\em both} 1/$e$ and
$1/(\sqrt2e)$. This has only one possible solution: $\kappa=0$. Thus,
the nonexistence of an AB effect arising
from the delta-function magnetic field
considered in \cite{barhel} might be used to eliminate charge
non-quantization, but it is equally true that with only a minor
change of logic the existence of charge non-quantization provides a
second reason for eliminating the fictitious delta-function magnetic
field.
Since neither reasoning seems advantageous over the other, we prefer
to invoke the physical reasoning outlined in our first objection
above, wherein the added
delta-function magnetic field is eliminated by making the standard
choice of gauge potential.

In summary,
unless there is an {\em a priori} reason
{\em independent of whatever charges are
present} for restricting the allowed values of the parameter
$\kappa$, the
derivation of a charge quantization condition by insisting that 
$\kappa$ be unobservable is firstly without teeth (since the charge
quantization condition so derived is so weak), and secondly, not even
necessary.
Furthermore, a much stronger objection can be raised, in that the
addition of an ad hoc
delta-function magnetic field, which is at the heart of
the charge quantization condition derived in \cite{barhel}, is
completely without physical motivation: if the discontinuous magnetic
field $B_{\rm sol}$ is to be avoided, one need merely smooth it out,
as would be the case in any case with a real solenoid. No fictitious
magnetic field arises, $\kappa=0$, and the quantization condition --
weak as it is -- never sees the light of day.

\bigskip

We thank Manu Paranjape for interesting conversations. This work was
funded in part by the
National Science and Engineering Research Council.

\end{document}